\documentclass[twocolumn,prl,showpacs,superscriptaddress]{revtex4}
\usepackage{amssymb}
\usepackage{amsmath}
\usepackage{graphicx}
\usepackage{subfigure}
\usepackage{natbib}
\usepackage{epsfig}
\usepackage{amsfonts}
\usepackage{mathrsfs}
\usepackage{CJK}
\usepackage[toc,page,title,titletoc,header]{appendix}

\begin{document}
\title{Classical-driving-assisted quantum speed-up}

\author{Ying-Jie Zhang}
 \affiliation{Shandong Provincial Key
Laboratory of Laser Polarization and Information Technology,
Department of Physics, Qufu Normal University, Qufu 273165, China}
 \affiliation{Beijing National Laboratory of Condensed Matter Physics, Institute of
Physics, Chinese Academy of Sciences, Beijing, 100190, China}

\author{Wei Han}
 \affiliation{Shandong Provincial Key
Laboratory of Laser Polarization and Information Technology,
Department of Physics, Qufu Normal University, Qufu 273165, China}

\author{Yun-Jie Xia}
 \affiliation{Shandong Provincial Key
Laboratory of Laser Polarization and Information Technology,
Department of Physics, Qufu Normal University, Qufu 273165, China}

\author{Jun-Peng Cao}
 \affiliation{Beijing National Laboratory of Condensed Matter Physics, Institute of Physics,
Chinese Academy of Sciences, Beijing, 100190, China}
\affiliation{Collaborative Innovation Center of Quantum Matter, Beijing, 100190, China}

\author{Heng Fan}
\email{hfan@iphy.ac.cn}
 \affiliation{Beijing National Laboratory of Condensed Matter Physics, Institute of Physics,
Chinese Academy of Sciences, Beijing, 100190, China}
\affiliation{Collaborative Innovation Center of Quantum Matter, Beijing, 100190, China}

\date{\today}
\begin{abstract}
We propose a method of accelerating the speed of evolution of an
open system by an external classical driving field
for a qubit in a zero-temperature structured reservoir. It is shown
that, with a judicious choice of the driving strength of the applied
classical field, a speed-up evolution of an open system can be
achieved in both the weak system-environment couplings and the
strong system-environment couplings. By considering the
relationship between non-Makovianity of environment and the
classical field, we can drive the open system from the Markovian to
the non-Markovian regime by manipulating the driving strength of
classical field. That is the intrinsic physical reason
that the classical field may induce the speed-up
process. In addition, the roles of
this classical field on the variation of quantum evolution speed in
the whole decoherence process is discussed.

\end{abstract}

\maketitle

{\it{Introduction.}}---In virtually all areas of quantum physics a
fundamental and important task arising is to drive a given initial state
to a target state in the minimum evolution time.
This problem involves in many areas of research such as quantum
communication \cite{01,1}, quantum metrology \cite{2}, quantum
computation \cite{3}, nonequilibrium thermodynamics \cite{03}, as
well as quantum optimal control protocols \cite{4,04,004,1004,0004}.
The minimum evolution time between two distinguishable states of a
system, which be defined as quantum speed limit time (QSLT)
\cite{5,05,6,06,006,0006,7,07,007,0007,1007,2007,8,08,9,10,11},
is a key method in characterizing the maximal speed of evolution of quantum
systems. Since the relevant influence of the environment on
processing or information transferring systems can not be ignored,
the unified bounds of evolution time including both Mandelstam-Tamm
(MT) and Margolus-Levitin (ML) types focused on the open system with
nonunitary dynamics process have been formulated \cite{9,10}.
Interestingly, the QSLT would equal to the actual driving time in
the weak system-environment couplings, while the strong
system-environment couplings can reduce the QSLT below the actual
driving time \cite{9}. This fact means that the strong
system-environment couplings can speed up the quantum evolution
process. However, under the weak system-environment couplings, the
accelerating of quantum evolution is generally not achieved without
any operating to the system. And as we all know that a speed-up
evolution of an open system would be preferable to deal with the
robustness of quantum simulators and computers against decoherence
\cite{011,0011}. So how to devise an effective and feasible
mechanism to speed up the evolution process of an open system under
more general physical conditions such as in weak-coupling case,
becomes extremely significant.

In this Letter, we will investigate a generic decoherence model of a
qubit interacting with a zero-temperature structured reservoir and
driven by an external classical field. We demonstrate how a speed-up
evolution of an open system can be acquired by manipulating the
driving strength of the classical field, although the
system-environment coupling is weak. By investigating the influence
of the classical field on the QSLT, for a certain critical driving
strength of the classical field, a sudden transition from no
speed-up to speed-up can occur in the weak-coupling regime.
Additionally under the strong-coupling regime, the speed of
evolution for the system can also be controlled to a speed-up or
speed-down process by the appropriate driving strength of the
classical field.

According to Ref. \cite{9}, the speed-up evolution of the open
system is mainly related to the non-Markovianity of the environment
\cite{012,12,0012,13,013,0013,014,0014,14,15}. So in order to clear
the physical reason of the speed-up process induced by the classical
field, we further focus on the relationship between the
non-Makovianity of environment and the classical field. We note that
the original Markovian dynamics can be changed to the non-Markovian
dynamics by choosing an agreeable driving strength of the classical
field. And the transition point from Markovian dynamics to
non-Markovian dynamics is interestingly equal to the critical driving strength
where the uniform evolution speed becomes the speed-up dynamical
process of the open system. Finally, we explore
the effects of the classical field on the variation of quantum
evolution speed in the whole decoherence process by calculating the
QSLT for the arbitrary time-evolution state. Remarkably, the applied
classical field can result in the smaller acceleration in the
speed-up process and the smaller deceleration in the speed-down
process.

{\it{Model.}}---Here, we consider a two-level system interacting
with a structured reservoir at zero temperature. A specific system
which consists of a two-level atom (transition frequency
$\omega_{0}$) interacting with a electromagnetic field has been
chosen in this Letter. And the atom is driven by a classical field with
frequency $\omega_{L}$. The Hamiltonian reads,
\begin{eqnarray}
H&=&\frac{\omega_{0}}{2}\sigma_{z}+\sum_{k}\omega_{k}a^{\dag}_{k}a_{k}+\Omega(e^{-i\omega_{L}t}\sigma_{+}+e^{i\omega_{L}t}\sigma_{-})\nonumber\\
&+&\sum_{k}g_{k}(a_{k}\sigma_{+}+a^{\dag}_{k}\sigma_{-}),\label{1}
\end{eqnarray}
where the operators $\sigma_{z}$ and $\sigma_{\pm}$ are defined by
$\sigma_{z}=|e\rangle{\langle}e|-|g\rangle{\langle}g|$,
$\sigma_{+}=|e\rangle{\langle}g|$, and
$\sigma_{-}=\sigma^{\dag}_{+}$ associated with the upper level
$|e\rangle$ and the lower level $|g\rangle$; $a_{k}$ and
$a^{\dag}_{k}$ are the annihilation and creation operators for the
field mode $k$, which is characterized by the frequency
$\omega_{k}$; $g_{k}$ and $\Omega$, both chosen to be real, are the
coupling constants of the interactions of the atom with the field
mode $k$ and with the classical driving field, respectively.
 In the
dressed-state basis
$\{|+\rangle=\frac{1}{\sqrt{2}}(|g\rangle+|e\rangle),|-\rangle=\frac{1}{\sqrt{2}}(|g\rangle-|e\rangle)\}$,
by considering two rotating reference frame through two unitary
transformation $U_{1}=\exp[-i\omega_{L}\sigma_{z}t/2]$ and
$U_{2}=\exp[i\omega_{0}\Sigma_{z}t/2]$ \cite{16,17}, the total
Hamiltonian in Eq. (\ref{1}) can be transferred to an effective
Hamiltonian in the rotating-wave approximation,
\begin{eqnarray}
H_{eff}=\frac{\omega'}{2}\Sigma_{z}+\sum_{k}\omega_{k}a^{\dag}_{k}a_{k}+\sum_{k}g'_{k}[a_{k}\Sigma_{+}+a^{\dag}_{k}\Sigma_{-}],\label{5}
\end{eqnarray}
with $\omega'=2\Omega+\omega_{0}$ and $g'_{k}=g_{k}/2$. Here $\Sigma_{z}$
and $\Sigma_{\pm}$ are defined by
$\Sigma_{z}=|+\rangle{\langle}+|-|-\rangle{\langle}-|$,
$\Sigma_{+}=|+\rangle{\langle}-|$, and
$\Sigma_{-}=\Sigma^{\dag}_{+}$. A noteworthy feature of this
effective Hamiltonian is that the basis states have been changed to
$\{|+\rangle,|-\rangle\}$, when the atom coupled with the structured
reservoir with the assistance of the external classical field.

Furthermore, at zero temperature, let us consider the situation
where the initial state of the system plus reservoir is of the form
$|\Psi(0)\rangle=|+\rangle_{S}|\mathbf{0}\rangle_{E}$, with
$|\mathbf{0}\rangle_{E}$ denotes the vacuum state of the reservoir.
By the Hamiltonian described in Eq. (\ref{5}), the
state of the total system at any time $t$ is given by, $
|\Psi(t)\rangle=c_{+}(t)|+\rangle_{S}|\mathbf{0}\rangle_{E}+\sum_{k}c_{k}(t)|-\rangle_{S}|1_{k}\rangle_{E}$,
where $|1_{k}\rangle_{E}$ is the state of the reservoir with only
one excitation in the $k$-th mode. The time evolution of the
probability amplitudes is governed by a series of differential
equations,
\begin{eqnarray}
\dot{c}_{+}(t)=-i\sum_{k}g'_{k}\exp[i(\omega'-\omega_{k})t]c_{k}(t),\\\label{7}
\dot{c}_{k}(t)=-ig'_{k}\exp[-i(\omega'-\omega_{k})t]c_{+}(t).\label{8}
\end{eqnarray}
Owing to no excitations in the initial state of the reservoir, that
is $c_{k}(0)=0$, we can obtain the integro-differential equation for
$c_{+}(t)$ as $
\dot{c}_{+}(t)=-\int^{t}_{0}dt_{1}f(t-t_{1})c_{+}(t_{1})$. The
correlation function $f(t-t_{1})$ is related to the spectral density
$S(\omega)$ of the reservoir. Here, the environment can be described
by an effective Lorentzian spectral density of the form $
S(\omega)=\frac{1}{2\pi}\frac{{\lambda}R}{(\omega-\omega_{c})^{2}+\lambda^{2}}$,
where $\lambda$ is the spectral width, $R$ the coupling strength,
and $\omega_{c}$ in the center frequency of the reservoir.
Typically, in weak-coupling regime ($\lambda>2R$), the behavior of
the qubit-cavity system is Markovian and irreversible decay occurs.
For strong-coupling regime ($\lambda<2R$), non-Markovian dynamics
occurs accompanied by an oscillatory reversible decay. Through
introducing the correlation function
$f(t-t_{1})={\int}d{\omega}S(\omega)e^{i(\omega'-\omega)(t-t_{1})}$
and performing the Laplace transform, we acquire
$s\tilde{c}_{+}(s)-c_{+}(0)=-\tilde{c}_{+}(s)\tilde{f}(s)$. From the
above equation we can derive the quantity $\tilde{c}_{+}(s)$.
Finally, inverting the Laplace transform we can obtains a formal
solution for the amplitude $c_{+}(t)=\varepsilon(t)c_{+}(0)$, with $
\varepsilon(t)=e^{-[\lambda-i(\omega'-\omega_{c})]t/2}[\cosh(Dt/2)+\frac{\lambda-i(\omega'-\omega_{c})}{D}\sinh(Dt/2)]$,
where
$D=\sqrt{\lambda^{2}-2R\lambda-(\omega'-\omega_{c})^{2}-2i(\omega'-\omega_{c})\lambda}$.
In the dressed-state basis, the reduced density matrix of the system
at time $t$ reads,
\begin{equation}
\rho_{t}=\left(
       \begin{array}{cccc}
        \rho_{++}(0)|\varepsilon(t)|^{2} & \rho_{+-}(0)\varepsilon(t) \\
         \rho_{-+}(0)\varepsilon^{*}(t) & 1- \rho_{++}(0)|\varepsilon(t)|^{2} \\
       \end{array}
     \right).\label{12}
\end{equation}

{\it{Speed-up of quantum evolution from $\rho_{0}$ to
$\rho_{\tau_{D}}$.}}---In order to illustrate the role of the
external classical field on the quantum speed of evolution of the
open system, we firstly start with the definition of the QSLT for
open quantum system. the QSLT can effectually define the bound of
minimal evolution time for arbitrary initial states, and be helpful
to analyze the maximal speed of evolution of open quantum system. A
unified lower bound, including both MT and ML types, has been
derived by Deffner and Lutz \cite{9}.
The QSLT is determined by an initial state $\rho_{0}=|\phi_{0}\rangle\langle\phi_{0}|$
and its target state $\rho_{\tau_{D}}$, governed by the master
equation $\dot{\rho}_{t}=L_{t}\rho_{t}$, with $L_{t}$ the positive
generator of the dynamical semigroup.
With the help of the von
Neumann trace inequality and the Cauchy-Schwarz inequality,
the QSLT is as follows,
\begin{equation}
\tau_{D}\geq\tau_{QSL}=\max\{\frac{1}{\Lambda^{1}_{\tau_{D}}},\frac{1}{\Lambda^{2}_{\tau_{D}}},\frac{1}{\Lambda^{\infty}_{\tau_{D}}}\}\sin^{2}[\mathfrak{B}(\rho_{0},\rho_{\tau_{D}})],
\label{13}
\end{equation}
with
$\Lambda^{p}_{\tau_{D}}=\tau_{D}^{-1}\int^{\tau_{D}}_{0}\|L_{t}\rho_{t}\|_{p}dt$,
and $\|A\|=(\sigma^{p}_{1}+\cdots+\sigma^{p}_{n})^{1/p}$ denotes the
Schatten $p$ norm, $\sigma_{1}$,$\cdots$,$\sigma_{n}$ are the
singular values of $A$,
$\mathfrak{B}(\rho_{0},\rho_{\tau_{D}})=\arccos\sqrt{\langle\phi_{0}|\rho_{\tau_{D}}|\phi_{0}\rangle}$
denotes the Bures angle between the initial and target states of the
quantum system. And the ML-type bound based on the operator norm
($p=\infty$) of the nonunitary generator provides the sharpest bound
on the QSLT \cite{9}. So in the following we will use this ML-type
bound to demonstrate the speed of the dynamics evolution from an
initial state $\rho_{0}$ to a final state $\rho_{\tau_{D}}$ by a
driving time $\tau_{D}$.
\begin{figure}[tbh]
\includegraphics*[bb=71 167 402 345,width=6.5cm, clip]{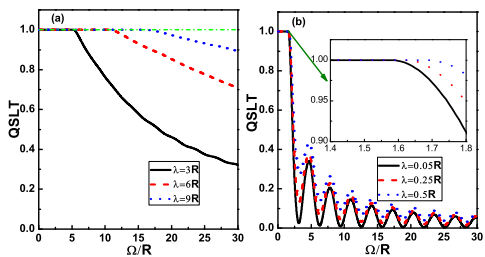}
\caption{(Color online) The QSLT for an open system driven by an
external classical field as a function of the parameter variable
$\Omega/R$, (a) in the weak-coupling regime ($\lambda>2R$), and (b)
in the strong-coupling regime ($\lambda<2R$). The green dash-dotted
line in (a) represents the actual driving time $\tau_{D}=1$.}
\end{figure}

We shall examine the dynamics process where the system starts in the
dressed state $|+\rangle$, that is $\rho_{++}(0)=1$ and
$\rho_{+-}(0)=0$. If there is no classical field to drive the system
($\Omega=0$), and the system is in resonance with the center mode of
the reservoir, i.e. $\omega_{0}=\omega_{c}$, the QSLT would equal to
the actual driving time $\tau_{D}$ in the weak-coupling regime
($\lambda>2R$), that is no speed-up dynamics process \cite{9}. In
order to obtain the quantum speed-up of the evolution process in the
weak-coupling regime, we show how to manipulate the QSLT for the
open system via a classical field. Fig. $1(a)$ shows the QSLT for an
open system as a function of the driving strength of the external
classical field $\Omega$ in the weak-coupling regime, with the
resonance case $\omega_{0}=\omega_{c}$ , and the actual driving time
$\tau_{D}=1$. It is worth noting that, a remarkable behavior of
sudden transition from no speed-up to speed-up can occur at a
certain critical driving strength of the classical field
$\Omega_{c}$. When $\Omega<\Omega_{c}$, the QSLT of the system is
actually the driving time, and then decreases monotonically with
increasing $\Omega$. So we therefore reach the interesting result
that the external classical field can be used to reduced the QSLT
below its value in the weak-coupling regime.
Thus we obtain the speed-up of the evolution of an open quantum system in the
weak-coupling regime. And then, numerical calculation also shows
that the critical driving strength $\Omega_{c}$ is determined by the
value of the spectral width $\lambda$. The larger the value of
$\lambda$ is, the lager the value of the critical driving strength
$\Omega_{c}$ should be requested. Take the cases in Fig. $1(a)$ as
examples, when $\lambda=3R$, we find the value of the critical
driving strength is $\Omega_{c}=5.31R$. While in the cases
$\lambda=6R$ and $\lambda=9R$, we can acquire $\Omega_{c}=10.89R$
and $\Omega_{c}=16.41R$, respectively.

Moreover, for the strong-coupling regime ($\lambda<2R$), the QSLT
exhibits a plateau independent of $\Omega$ for the moderate driving
strength of the classical field, and then periodically decrease for
the large driving strength of the classical field (as shown in Fig.
$1(b)$). That is to say, in the strong-coupling regime, the speed of
evolution for the system can be controlled to a speed-up or
speed-down process by the driving strength $\Omega$.

As noted in Refs. \cite{9,10,11}, the non-Markovianity in the
dynamics process ($\rho_{0}$ to $\rho_{\tau_{D}}$), and the
associated information backflow from the reservoir, can lead to
faster quantum evolution, and hence to smaller QSLT. In order to
understand the physical reason of the speed-up process, in what
follows we would describe a scheme how to turn the dynamics from
Markovian to non-Markovian by adding an external classical field to
the qubit.
\begin{figure}[tbh]
\includegraphics*[bb=37 87 570 500,width=6.5cm, clip]{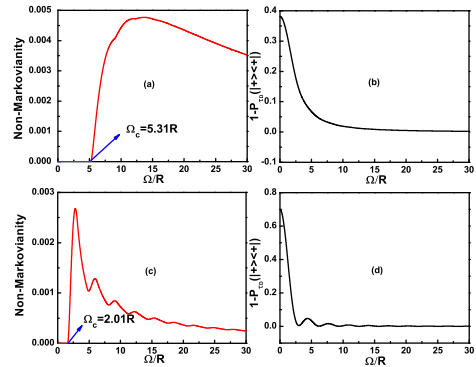}
\caption{(Color online) The non-Markovianity $\mathbb{N}(\Phi)$ (red
solid line) and the population $1-P_{\tau_{D}}(|+\rangle\langle+|)$
(black solid line) for an open system driven by an external
classical field as a function of the parameter variable $\Omega/R$,
with $\tau_{D}=1$. (a) and (b) in the weak-coupling regime,
$\lambda=3R$, (c) and (d) in the strong-coupling regime,
$\lambda=0.05R$. }
\end{figure}
The measure $\mathbb{N}(\Phi)$ for non-Markovianity of the quantum
process $\Phi(t)$ has been defined by Breuer $et$ $al.$ \cite{12}.
Considering a quantum process $\Phi(t)$, $\rho(t)=\Phi(t)\rho(0)$,
where $\rho(0)$ and $\rho(t)$ denote the density operators at time
$t=0$ and at any time $t>0$, respectively, then the non-Markovianity
$\mathbb{N}(\Phi)$ is defined as
$\mathbb{N}(\Phi)=\max_{\rho_{1,2}(0)}\int_{\sigma>0}dt\sigma(t,\rho_{1,2}(0))$,
where $\sigma(t,\rho_{1,2}(0))$ is the rate of change of the trace
distance,
$\sigma(t,\rho_{1,2}(0))=\frac{d}{dt}\mathcal{D}(\rho_{1}(t),\rho_{2}(t))$.
The trace distance $\mathcal{D}$ describing the distinguishability
between the two states is defined as \cite{01}
$\mathcal{D}(\rho_{1},\rho_{2})=\frac{1}{2}\|\rho_{1}-\rho_{2}\|$,
where $\|M\|=\sqrt{M^{\dag}M}$ and $0{\leq}\mathcal{D}\leq1$. And
$\sigma(t,\rho_{1,2}(0))\leq0$ corresponds to all dynamical
semigroups and all time-dependent Markovian processes, a process is
non-Markovian if there exists a pair of initial states and at
certain time $t$ such that $\sigma(t,\rho_{1,2}(0))>0$. We should
take the maximum over all initial states $\rho_{1,2}(0)$ to
calculate the degree of non-Markovianity. Similar to Refs.
\cite{12,13}, by drawing a sufficiently large sample of random pairs
of initial states, the optimal state pair is attained for the
initial states $\rho_{-}(0)=|-\rangle{\langle}-|$ and
$\rho_{+}(0)=|+\rangle{\langle}+|$ by strong numerical calculations
in the dressed-state basis $\{|-\rangle,|+\rangle\}$. Here, for the
optimal state pair, the rate of change of the trace distance can be
acquired $\sigma(t,\rho_{+,-}(0))=\partial_{t}|\varepsilon(t)|^{2}$,
and the singular values of the nonunitary generator $L_{t}\rho_{t}$
are given by $|\sigma(t,\rho_{+,-}(0))|$.

In the light of Eq. (\ref{13}), the QSLT for the system can be
clearly derived as $
\tau_{QSL}=\frac{\tau_{D}[1-P_{\tau_{D}}(|+\rangle\langle+|)]}{\int^{\tau_{D}}_{0}|\sigma(t,\rho_{+,-}(0))|dt}
=\frac{\tau_{D}[1-P_{\tau_{D}}(|+\rangle\langle+|)]}{2\mathbb{N}(\Phi)+1-P_{\tau_{D}}(|+\rangle\langle+|)},$
where $P_{\tau_{D}}(|+\rangle\langle+|)=|\varepsilon(\tau_{D})|^{2}$
is the population of the dressed-state $|+\rangle$ at time
$\tau_{D}$. It is easy to find that the QSLT is strictly related to
the non-Markovianity of the evolution from $\rho_{0}$ to
$\rho_{\tau_{D}}$ and the population of the dressed-state
$|+\rangle$ at time $\tau_{D}$. Then we investigate the effects of
the driving classical field on the non-Markovianity
$\mathbb{N}(\Phi)$ and the population
$1-P_{\tau_{D}}(|+\rangle\langle+|)$, as shown in Fig. $2$. Both
in the weak-coupling regime ($\lambda=3R$) and in the strong-coupling
regime ($\lambda=0.05R$), by considering the evolution process
within the driving time, Figs. $2(a)$ and $2(c)$ illustrate that the
original Markovian dynamics can be changed to the non-Markovian
dynamics by choosing an agreeable driving strength of the classical
field. And the transition point from Markovian dynamics to
non-Markovian dynamics is equal to the critical driving strength of
the classical field $\Omega_{c}$ where the uniform evolution speed
becomes the speed-up dynamical process of the system. On the other
hand, a nonmonotonic behavior of the non-Markovianity can be shown
in Fig. $2$: when $\Omega>\Omega_{c}$, the non-Markovianity firstly
increase with increasing $\Omega$, after it reaches a maximum value,
it decreases with further increasing of $\Omega$. However, Figs.
$2(b)$ and $2(d)$ show that the population
$1-P_{\tau_{D}}(|+\rangle\langle+|)$ converges to zero for the
strong driving classical field in the weak-coupling regime (no
oscillations) and the strong-coupling regime (oscillations are
present). So the QSLT can still be reduced by the classical field,
that means the external classical field can be used to control the
speed of evolution of quantum systems.

 {\it{Variation of quantum evolution speed of the whole decoherence process.}}---One may naturally concern the variation of a speed for QSLT based on
an arbitrary time-evolution state $\rho_{\tau}$. The QSLT for mixed
initial states \cite{10} can be used to demonstrate the quantum
speed of evolution from $\rho_{\tau}$ to another
$\rho_{\tau+\tau_{D}}$ by a driving time $\tau_{D}$. Here, we mainly
examine the whole dynamics process where the system starts from the
dressed state $|+\rangle$ in the weak-coupling regime. By
calculating the singular values of $\rho_{\tau}$ and
$L_{t}\rho_{t}$, the singular values for $\rho_{\tau}$ are
$\varrho_{1}=P_{\tau}(|+\rangle\langle+|)$ and
$\varrho_{2}=1-P_{\tau}(|+\rangle\langle+|)$, while for
$L_{t}\rho_{t}$, the singular values are
$\sigma_{1}=\sigma_{2}=|\dot{P}_{t}(|+\rangle\langle+|)|$.
\begin{figure}[tbh]
\includegraphics*[bb=44 196 500 370,width=8.5cm, clip]{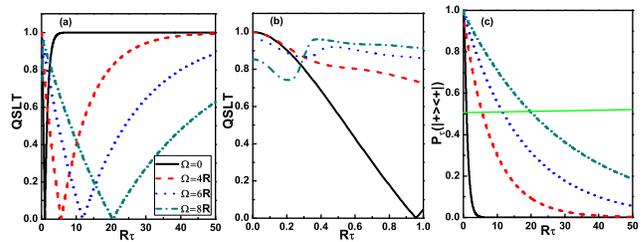}
\caption{(Color online) The QSLTs (a), (b) and the population
$P_{\tau}(|+\rangle\langle+|)$ (c) for the arbitrary time-evolution
state as a function of the initial time parameter $R\tau$ by
different driving strength $\Omega$ of the applied classical field.
Here we consider an open dynamics process started from the dressed
state $|+\rangle$, and parameters are chosen as $\lambda=3R$,
$\tau_{D}=1$. The green dash-dot-dotted line in (c) represents the
population $P_{\tau_{c}}(|+\rangle\langle+|)=0.5$, which can be used
to determine the critical time point between the speeded-up process
and the speed-down process.}
\end{figure}
Then the QSLT for a time-evolution state $\rho_{\tau}$ can be
calculated
$\tau_{QSL}=\frac{\tau_{D}|[1-2P_{\tau}(|+\rangle\langle+|)][P_{\tau}(|+\rangle\langle+|)-P_{\tau+\tau_{D}}(|+\rangle\langle+|)]|}{\int^{\tau+\tau_{D}}_{\tau}|\dot{P}_{t}(|+\rangle\langle+|)|dt}.$
Figs. $3(a)$ and $3(b)$ present the results of our analysis for
$\tau_{QSL}$ by choosing different driving strength $\Omega$ of the
applied classical field, with $\lambda=3R$. By adding an external
classical field to the system, we observe that, the evolution of the
open system can first execute a speed-up process and then show
gradual deceleration process in the case $\Omega<\Omega_{c}=5.31R$.
The case $\Omega>\Omega_{c}=5.31R$ is complicated which can be
explained by non-Markovianity, see Fig. $3(b)$. Overall, a
remarkable result we find that, for the speed-up process, the decay
rate of the QSLT can decrease with the driving strength $\Omega$
increasing. And for the speed-down process, the increasing rate of
the QSLT would also be reduced by choosing a stronger driving
classical field. This can be understood that the applied classical
field can lead to the smaller acceleration in the speed-up process,
and also the smaller deceleration in the speed-down process. This is
a newly noticed phenomenon. Finally, as shown in Fig. $3(c)$, the
increasing of the driving strength $\Omega$ makes the energy
exchange between the system and the environment more slow. This
behavior plays the dominating role on the variation of quantum
evolution speed in the whole decoherence process.

{\it{Conclusion.}}---In summary, we demonstrated that a speed-up
evolution of an open system could be achieved by manipulating the
driving strength of an external classical field. We show that
the phenomenon of transition from Markovian to non-Markovian dynamics induced by the
classical field is the main physical reason of the speed-up process.
Recent experiments with photons allow one to drive the open
system from Markovian to non-Markovian regime in the
dephasing channels \cite{14,15}. In comparison, the
results we illustrated here involve the amplitude
damping channels for the controlling of Markovian and
non-Markovian dynamics.
The potential candidates which can realize this type of quantum state
and the environment can be systems such as cavity QED \cite{18}, trapped ions \cite{19},
superconducting qubits \cite{you1} and
the Nitrogen-Vacancy center of diamond \cite{20}.

\emph{Acknowledgements.}---This work was supported by ¡°973¡±
program under grant No. 2010CB922904, the National Natural Science
Foundation of China under grant Nos. 11304179, 11247240, 11175248,
61178012, the Specialized Research Fund for the Doctoral Program of
Higher Education under grant Nos. 20133705110001, 20123705120002,
the Provincial Natural Science Foundation of Shandong under grant
No. ZR2012FQ024, and the Scientific Research Foundation of Qufu
Normal University.

\end{document}